\documentstyle[prb,aps,floats,epsf,preprint]{revtex}
\begin{document}
\draft
\title{The Gibbs-Thomson formula at small island sizes - corrections 
for high vapour densities}
\author{Badrinarayan Krishnamachari, James McLean \cite{add}, Barbara Cooper,
James Sethna}
\address{Laboratory of Atomic and Solid State Physics, Cornell
University, Ithaca, NY 14853}
\date{\today}
\maketitle

\begin{abstract}
In this paper we report simulation studies of equilibrium features,
namely circular islands on model surfaces, using Monte-Carlo methods.
In particular, we are interested in studying the relationship between
the density of vapour around a curved island and its curvature.  The
``classical'' form of this relationship is the Gibbs-Thomson formula,
which assumes the vapour surrounding the island to be an ideal gas.
Numerical simulations of a lattice gas model, performed for various
sizes of islands, don't fit very well to the Gibbs-Thomson formula.
We show how corrections to this form arise at high vapour densities,
wherein a knowledge of the exact equation of state (as opposed to the
ideal gas approximation) is necessary to predict this relationship.
By exploiting a mapping of the lattice gas to the Ising model one can
compute the corrections to the Gibbs-Thomson formula using high field
series expansions. The corrected Gibbs-Thomson formula matches very
well with the Monte Carlo data.  We also investigate finite size
effects on the stability of the islands both theoretically and through
simulations.  Finally the simulations are used to study the
microscopic origins of the Gibbs-Thomson formula.  It is found that
smaller islands have a greater adatom detachment rate per unit length
of island perimeter.  This is principally due to a lower coordination
of edge atoms and a greater availability of detachment moves relative
to edge moves.  A heuristic argument is suggested in which these
effects are partially attributed to geometric constraints on the
island edge.
\end{abstract}


\section{Introduction}
\label{sec:intro}
The study of the stability and evolution of nanoscale features is
useful in understanding microscopic processes involved in the
formation and growth of solids. Theoretical studies of the coarsening
of an ensemble of ``islands'' \cite{ostwald} as well as models for the
decay of single nanoscale ``islands'' \cite{peale,isldecay,rosenfeld},
make use of the fact that there exists a high vapour pressure in
equilibrium with extremely small islands on the surface.  These
theories which describe systems away from equilibrium make use of the
relationship between the equilibrium vapour pressure around a circular
island and the curvature of the island, which is given by the
Gibbs-Thomson formula.  In this paper we shall take a closer look at
this formula and show that it needs important corrections at high
vapour densities wherein interaction between atoms of the vapour
cannot be ignored.  We will discuss the two dimensional problem of an
island in equilibrium with a vapour of adatoms on the surrounding
terrace.  We will ignore the (often small) three dimensional bulk
evaporation-condensation and bulk vapour pressure.

For a two dimensional island of radius $r$ in equilibrium with the
vapour of adatoms around it, the Gibbs-Thomson formula
\cite{abr,widom} is 
\begin{equation}
p(r) = p_{\infty}\exp\left( \gamma/\left( r \rho_{s} k T \right) \right),
\label{gtpress}
\end{equation}
where $p_{\infty}$ is the vapour pressure outside a straight interface
between solid and vapour, $\gamma$ is the edge free energy per unit
length of the two dimensional island on the substrate, $\rho_{s}$ is
the density of the solid island, $k$ is Boltzmann's constant and $T$
the absolute temperature.  This relation assumes that the gas
surrounding an island is ``ideal'' and hence we may write down a
similar expression for the density of the gas in equilibrium with an
island of radius $r$ as
\begin{equation}
\rho(r) = \rho_{\infty}\exp\left( \gamma/\left( r \rho_{s} k T \right) \right),
\label{gt}
\end{equation}
The above equation is often seen in the context of nucleation theory
of growth in first order phase transformations \cite{abr} in addition
to its application to the study of equilibrium and decay of features
on surfaces.

Section \ref{sec:gt} discusses the derivation of the ``classical''
Gibbs-Thomson formula for a finite size system having a constant
number of atoms.  We simulate a two-dimensional lattice gas on a
square lattice, using Monte Carlo techniques, in order to test this
relation and find that the Gibbs-Thomson formula deviates
significantly from the data from our simulation (section
\ref{sec:sim}). This is because of the assumption that the vapour
around the island is an ideal gas. In our case, we can map the lattice
gas to the Ising model, enabling us to use high field series
expansions to generate an equation of state for the lattice gas that
improves upon the ideal gas assumption.  This is used to derive a
corrected Gibbs-Thomson formula in section \ref{sec:ising}. This
corrected Gibbs-Thomson formula gives a very good description of the
data obtained from the simulation. In section \ref{sec:stab} we
discuss the constraint of finite size along with predictions regarding
the stability of the islands. We investigate the microscopic origins
of the enhanced vapour pressure around small islands in section
\ref{sec:mic} and present a plausible argument in which we try to
correlate the enhancement with geometric constraints on the island.  We
finally conclude with section \ref{sec:conc}.

\section{The Gibbs-Thomson formula} 
\label{sec:gt}
The Gibbs-Thomson formula is encountered frequently in the study of
curved interfaces in equilibrium\cite{widom}.  It is also encountered
in the context of nucleation and critical droplet theory (for first
order phase transformations)\cite{abr} wherein one studies the
formation of droplets of liquid (analogous to the solid islands
mentioned in the introduction) in a supersaturated gas and the free
energy barrier to the formation of these droplets. However, in this
context, the droplet formed is often at a saddle point of the total
free energy of the system, in short an unstable, stationary state.
These droplets can be stabilized by finite size effects \cite{jim}. If
the system under study (with a fixed number of atoms) is placed in
a box of fixed volume and temperature then one can show that under
certain conditions the global minimum of the free energy of the system
consists of a droplet/island in equilibrium with its vapour and
the relationship between the island size and vapour pressure is
given by the Gibbs-Thomson formula.

We will now derive the Gibbs-Thomson formula for this system.
Consider $N$ atoms of supersaturated vapour in a two dimensional box of
volume $V$, at a temperature $T$. The system is at a metastable state
on its phase diagram (point 1 in Fig. \ref{pdvfig}), because the
supersaturated vapour can lower its Helmholtz free energy by
nucleating a solid island (point 5 on the phase diagram), which would
be in equilibrium with the remaining vapour around it (point 2)
\cite{foot1}.  We will show this explicitly by computing the change in
free energy of the system upon nucleation of an island.

The change in Helmholtz free energy of the system on nucleating a
solid island of radius $r$, from the supersaturated vapour, has three
pieces to it:
\newline 
a) An increase in edge free energy of the
island formed given by
\begin{equation} 
\Delta F_{\text{edge}} = 2 \pi r \gamma, 
\end{equation}
where $\gamma$ is the line tension or free energy per unit length of
the edge.
\newline
b) A change in the bulk free energy of the condensing atoms. If the
number density of the solid formed is $\rho_{s}$, the decrease in free
energy is computed by considering the free energy changes along the
isotherm 1-2-3-4-5 in Fig. \ref{pdvfig} and works out to be 
\begin{equation}
\Delta F_{c} = \rho_{s} \pi r^{2}kT \ln \left( \frac{\rho_{\infty}}{\rho_{i}} \right) 
- \pi r^{2} kT (\rho_{\infty} - \rho_{s}).
\end{equation}
Here $\rho_{\infty}$ is the number density of the gas when it is in
equilibrium with a straight interface at point 3 of the phase diagram
and $\rho_{i} \equiv N/V$ the initial number density of the vapour.
The free energy changes are computed by integrating the differential
change in free energy at constant temperature, $dF = -pdV$.  The first
term represents the change in free energy along path 1-2-3 assuming
the supersaturated vapour to behave as an ideal gas and the second
term represents the free energy change along path 3-4. We have
neglected the change in free energy of the solid when it is compressed
to a high pressure along path 4-5.  This is equivalent to assuming
zero compressibility for the solid phase.  In most physical situations
even though the compressibility of the solid phase is not exactly
zero, the slope of the isotherm on the P-V curve is very high.
Consequently the corresponding contribution to the free energy change
is small and the assumption we make is therefore reasonable.  We have
also derived the Gibbs-Thomson formula with a non-zero compressibility
for the solid by assuming the vacancies in the solid to behave as an
ideal gas.  However we do not describe this here.  The results from
such an assumption produce an imperceptible change in the plots of the
Gibbs-Thomson formula at the densities and temperatures of interest to
us.
\newline
c) A decrease in free energy of the non-condensing atoms as they
expand to occupy the region left vacant by the condensing atoms, 
\begin{equation}
\Delta F_{\text{nc}} = - \left(N - \rho_{s} \pi r^{2}\right) k T
\ln \left( {\frac{V-\pi r^{2}}{V-\rho_{s} \pi r^{2}V/N}} \right).
\end{equation}
The total free energy change is the sum of the above three pieces 
\begin{equation}
\Delta F_{\text{tot}} = 2 \pi r \gamma +  \rho_{s} \pi r^{2}kT \ln \left( \frac{\rho_{\infty}}{\rho_{i}} 
\right) - \pi r^{2} kT (\rho_{\infty} - \rho_{s}) - \left(N - \rho_{s} \pi r^{2}\right) k T
\ln \left( {\frac{V-\pi r^{2}}{V-\rho_{s} \pi r^{2}V/N}} \right).
\label{dftot}
\end{equation}
This is plotted for $\rho_{s}=0.996$, $T=1347 K$,
$\rho_{\infty}=0.0036$, $\gamma = 0.1173$, $N=150$, $V=10,000$ in
Fig. \ref{dffig}.  This choice of numbers will become clear in
sections \ref{sec:sim} and \ref{sec:ising} where we describe
simulations performed with these parameters. It can be seen from
Fig. \ref{dffig} that the free energy has four extrema: a minimum (I) at
which an island is in true equilibrium with its surrounding vapour; a
maximum (U), at which a smaller island is in metastable
equilibrium with the surrounding vapour; the unstable vapour phase
itself (V) and the unstable solid phase (S).  Extremizing the total 
free energy w.r.t. $r$ yields
\begin{equation}
\ln \left( \frac{\rho_{f}}{\rho_{\infty}} \right) = 
\frac{\gamma}{r \rho_{s} k T} + \frac{\rho_{f} - \rho_{\infty}}{\rho_{s}},
\label{consgt}
\end{equation}
where $\rho_{f} \equiv (N - \rho_s{s} \pi r^{2})/(V - \pi r^{2})$,  
is the number density of the vapour surrounding the
island.  This form for the relationship between the radius of the
island and the density of vapour surrounding it is true at both the
maximum (U) and the minimum (I) and yields two roots for $r$ at
constant $N$ and $V$, only one of which is stable.  The second term on
the right hand side of Eq. (\ref{consgt}) is usually small\cite{widom}
and is often neglected to yield a form for the density which is
identical to Eq. (\ref{gt}).  This approximation is justified in our
case too; a point we shall return to at the end of the next section.

\section{Simulation details}
\label{sec:sim}
We perform Monte-Carlo simulations of a lattice gas of ``atoms''
constrained to a single layer.  The lattice gas Hamiltonian (for a
square lattice in two dimensions) can be written as
\begin{equation}
{\cal H_{G}}= -\epsilon \sum_{<i, j>}{n_{i} n_{j}},
\label{hgas}
\end{equation}
where $n_{i} = 1$ \ or \ $0$ depending on whether site $i$ is occupied
by an atom.  The sum runs over nearest neighbour ($<i,j>$) pairs
and reduces the total energy by $-\epsilon$ whenever two nearest
neighbour sites are occupied. Thus $\epsilon$ represents a bond
energy.  We now briefly describe details of the simulation.

We use a continuous time Monte Carlo (MC) scheme \cite{kalos} that
helps reduce the time required to run the simulations.  Barriers for
moves of atoms in the MC were based on barriers for the Cu (100)
surface calculated using effective medium theory \cite {barriers}.
They are allowed to depend on the coordination of the atom both 
before and after it makes a move. The barriers used are shown
in Table \ref{bartab}.  The barriers are not all independent since
they satisfy the constraint of detailed balance. Details regarding the
choice of barriers as well as the number of barriers can be found in
the paper referring to decay of these island like features
\cite{isldecay}, along with some other details regarding the simulation.
The choice of barriers cannot affect the macroscopic static
equilibrium behaviour of the islands, but definitely plays a role in
its dynamics.  Macroscopic static behaviour in equilibrium is governed
solely by the bond energy. This is chosen to be $\epsilon = 0.341$
eV. For this bond energy, the critical temperature (at which all solid
melts into gas) is $T_{c}$ = 2245 K. This is known from the critical
temperature of the Ising model to which this model can be mapped, as
described later on in this section. Simulations were performed at
temperatures of 1347 K and 1000 K, both well below the critical
temperature.  The system size was 100x100 lattice units and we ran the
simulation by letting islands of different sizes come to equilibrium
with their vapour.  Time scales are governed by a global attempt
frequency which was set to $\nu = 10^{12} s^{-1}$.  The initial
configuration in each run was a circular island, with no adatoms around
it, sitting at the centre of a vacant terrace, with periodic boundary
conditions.  The island would quickly source out atoms onto the
terrace and come to equilibrium with this gas of atoms. The
equilibrium between island and vapour is signalled by an island whose
size fluctuates in time around a stable mean value. Fig. \ref{pictfig}
shows a snapshot of one of these islands in equilibrium with its
vapour as seen in the simulation. Typically each of these runs made
about 40 million to a 100 million MC moves and took about 4 to 9 hours
of CPU time on a IBM RS6000.

Once the island has come to equilibrium with its vapour one can
compute its radius from a knowledge of its average size and one can
also compute the average density of the gas around the island, by
averaging at regular intervals of time, uncorrelated reports of the
density. This is done for each of the islands of different initial
size that we ran at the two temperatures mentioned above. There are
various definitions possible for the radius of an island
\cite{widom}. We compute its radius using the relation $area = \pi
r^{2}$, where the area can be computed from the snapshots of the
island that are reported (it includes the area of vacancies inside the
island).  The radius thus computed is equivalent to the equimolar
radius $r_{e}$ defined by Gibbs \cite{widom}. All length scales are
measured in units of the lattice spacing which is set to 1.

The density of the gas is computed by counting the number of atoms on
the terrace and then dividing this by the area of the terrace that is
free for occupation by the gas. Care is taken to exclude a one-lattice
spacing zone around the island as this cannot be occupied by an atom
of the vapour (if it were it would be part of the island).  In order
to perform statistics we first compute the correlation time for the
data.  This is done by computing the autocorrelation of the island
size as a function of time (in equilibrium).  Typically the
autocorrelation decays with some time constant $\tau$.  We then
consider data points which are separated by more than a couple of time
constants, as independent in time.  Essentially we bin the data into
bins of size about $2\tau$ replacing the data with its average value
in each bin.  We then take an average of these average values and
compute the standard deviation assuming the average data point in each
bin to be uncorrelated with that in xsanother bin.  The same procedure
is adopted to determine the density of gas around the island.  This is
how the error bars are obtained for plotting purposes.

Fig. \ref{gtfig} shows a plot of the logarithm of the density
vapour versus the curvature ($1/r$) of the island, for the two
different temperatures.  In order to compare the data to the
prediction from the Gibbs-Thomson formula [Eq. (\ref{gt})] we need the
edge free energy $\gamma$, the density of the solid deep inside the
bulk $\rho_{s}$, and the density of the vapour outside a straight
interface $\rho_{\infty}$.  These can be obtained by exploiting a
mapping of the lattice gas to the Ising model, outlined below.

The Hamiltonian for the lattice gas [Eq. (\ref{hgas})] can be made to
resemble that of an Ising model, using the transformation $n_{i} =
(1+s_{i})/2$, to give
\begin{equation}
{\cal H_{I}} = -\epsilon/4 \sum_{<n.n>}{s_{i} s_{j}} - \epsilon 
\sum_{i}{s_{i}} - N\epsilon/2,
\label{hising}
\end{equation}
where $N$ is the total number of sites on the lattice and the spin $s_{i}$ 
takes on values of $\pm 1$.  The second term would be analogous to a field
term in the Ising model with an external field of strength $\epsilon$.

This mapping helps us determine the parameters $\gamma$,
$\rho_{\infty}$ and $\rho_{s}$, that are relevant to this simulation.
The edge free energy (i.e., surface tension), $\gamma$, is known as a
function of temperature and orientation of the normal to the surface
for the case of the 2 dimensional Ising model \cite{zia}.  It varies
between a maximum and minimum value indicated in Table \ref{isingtab}
and we see that the variation is not significant at the two
temperatures at which we perform the simulations.  We use an average
value for the surface tension which we approximate as
\begin{equation}
\gamma_{\text{avg}} =  \frac{\int \gamma ds}{\int ds} 
\approx \frac{\int \gamma^{2}d\theta}{\int \gamma d\theta}.
\end{equation}
The results of averaging are also indicated in Table \ref{isingtab}.
Once again note that length scales are measured in terms of the
lattice spacing which is set to 1.  The values for $\rho_{\infty}$ and
$\rho_{s}$ are known from the spontaneous magnetization.  Using the
mapping for lattice gas to Ising variables these can be calculated as
$\rho_{\infty} = (1 - m)/2$ and $\rho_{s} = (1+m)/2$, where $m$ is the
spontaneous magnetization. The values of $\rho_{\infty}$ and
$\rho_{s}$ are also indicated in Table \ref{isingtab}.  Note that the
density of the solid $\rho_{s}$ is not identically equal to one.  This
is because of the presence of vacancies inside the solid, which can
be seen even in the simulation.  With this we have the three
parameters necessary to plot the Gibbs-Thomson formula.

The dashed line in Fig. \ref{gtfig} is the ``classical'' Gibbs-Thomson
prediction for the relationship between the density of vapour and
radius of the island as defined in [Eq. (\ref{gt})].  We see that the
formula is satisfactory at large radii and low temperatures but
important corrections are needed elsewhere.  The next section
discusses corrections to the ``ideal-gas'' equation of state used in
the derivation of the Gibbs-Thomson formula \cite{foot2} .  Note one
may just fit the data to an exponential form given by the
Gibbs-Thomson formula.  This yields a value for the surface
tension of $1.59\gamma_{\text avg}$.  As one can see this
is 60 percent off from the average value one would expect from the
Ising model results.  However this is useful in fitting the data to an
analytic expression of the Gibbs-Thomson form with a pre-factor in the
exponent viz.,  
$\rho_{\infty}\exp(\alpha\gamma/(r\rho_{s}kT))$, where $\alpha = 1.59$.

\section{Corrected Gibbs-Thomson formula for the Ising Model} 
\label{sec:ising}
The mapping from the lattice gas to the Ising model was discussed in
Section \ref{sec:sim}.  This enables us to compute properties of the
lattice gas system from a knowledge of the corresponding Ising system.
We will be interested in obtaining corrections to the Gibbs-Thomson
formula that take into account the ``non-ideal'' nature of the gas of
adatoms surrounding an island.  To this end we rederive the
Gibbs-Thomson formula using a more accurate equation of state than the
ideal gas one for the lattice gas/Ising system, using high field
series expansions.

\begin{figure}
\epsfxsize=8.6 cm
\epsffile{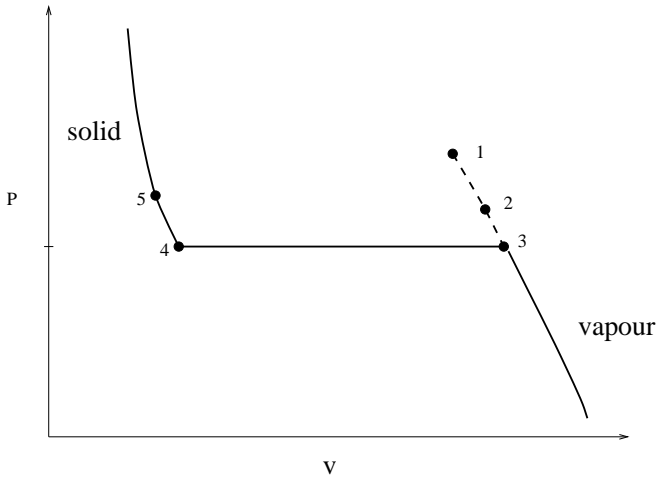}
\caption{Equation of state for the ideal gas}
\label{pdvfig}
\end{figure}

\begin{figure}
\epsfxsize=8.6 cm
\epsffile{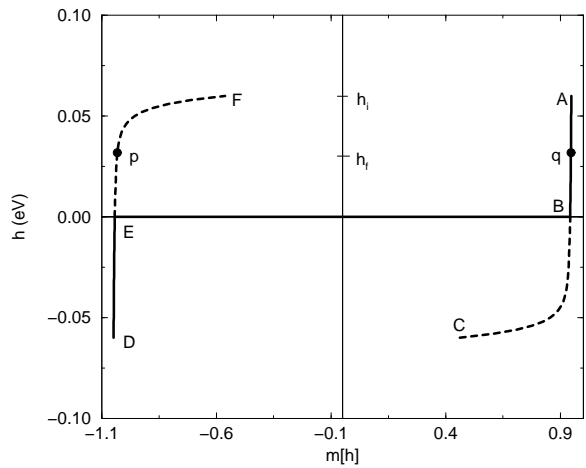}
\caption{Equation of state for the Ising Model}
\label{mvshfig}
\end{figure}

One can obtain the Helmholtz free-energy per site of the Ising model
(as a function of field, at a fixed temperature) by means of series
expansions, starting from a very high value of the field.  The first 4
terms of such an expansion of the equilibrium free energy for $h>0$
are,
\begin{eqnarray}
f^{>}[h] & = & -h - \epsilon/2 - kt(\omega x^{4} + \omega^{2}(2x^6 -
2.5x^{8}) +
\omega^{3}(6x^{8} - 16x^{10} +31/3x^{12}) \nonumber \\
     & &\mbox{} + \omega^{4}(x^{8} + 18x^{10} - 85x^{12} + 118x^{14} -
209/4x^{16}) + \dots),
\label{fseries}
\end{eqnarray}
where $\omega \equiv \exp(-2h/(kT))$, $x \equiv
\exp(-\epsilon/(2kT))$, $k$ is Boltzmann's constant and $T$ the
absolute temperature.  The coefficients of various terms in this
expansion are obtained analogous to low temperature expansions
\cite{parisi,essam}.  We use the first thirteen terms of this
expansion in our analysis. Differentiating the above expansion w.r.t
field yields an expansion for the magnetization per-site as a function
of field, for $h > 0$.  The magnetization is odd in $h$ (note the
expansion isn't),
\begin{eqnarray}
m^{>}[h] & = & 1 - 2(\omega x^{4} + 2\omega^{2}(2x^{6} - 2.5x^{8}) +
3\omega^{3}(6 x^{8} - 16x^{10} + 31/3x^{12}) \nonumber \\ & & \mbox{}
+ 4\omega^{4}(x^{8} + 18x^{10} - 85x^{12} + 118x^{14} - 209/4x^{16}) + \dots).
\label{mseries}
\end{eqnarray}
The expressions for f[h] and m[h] for $h < 0$ can be obtained by using
the up-down symmetry of the Ising model. Thus $f^{<}[h] = f^{>}[-h]$
for $h < 0$ and $m^{<}[h] = -m^{>}[-h]$ for $h < 0$.  This can be used
to plot the equation of state for this system (Fig. \ref{mvshfig}).
For large positive values of the field the state is essentially one in
which all the spins are pointing up (or all $n = 1$, the solid phase)
Conversely, the spins are all pointing down (gaseous phase of adatoms)
for large negative values of the field. The dashed portions BC and EF
on the equation of state represent metastable states and are analytic
continuations of the equilibrium equation of state $m[h]$, i.e., we
use $m^{>}[h]$ as given by Eq. (\ref{mseries}) for $h < 0$ to generate
the curve BC, on the equation of state.  Note the similarity between
this equation of state and the equation of state for an ideal gas
(Fig. \ref{pdvfig}).  Adatoms and solid can co-exist in equilibrium at
zero field.  In this case one has a flat interface between solid and
gas. In addition to this one could have metastable states of the
system wherein adatoms and solid co-exist at a finite field (e.g.,
states p and q on the equation of state co-exist at a field value of
$h_{f}$).  However, in this case one could have a solid with a finite
radius of curvature (just as in the ideal gas case: points 5 and 2 in
Fig. \ref{pdvfig}).  In order to compute the radius of the solid in
equilibrium with the gas of adatoms around it one can compute the free
energy change in nucleating a solid, in a system of pure gas which is
at state F on the phase diagram. The procedure adopted is similar to
the one in Section \ref{sec:gt}.  However, one has to minimize the
appropriate thermodynamic potential.  For the ordinary Ising model
(non-conserved order parameter) the Helmholtz free energy is at a
minimum in the equilibrium state at constant temperature, volume and
external field.  Since we work with a constant number of atoms in the
lattice gas, the total magnetization of the Ising model is held fixed
($ M \equiv \sum_{i} s_{i} = $ const.). Consequently one would have to
minimize the Legendre transform of the Helmholtz free energy, which we
shall henceforth refer to as the free energy, $G(T, V, M) = F + Mh$,
(it could also be called a thermodynamic potential). Consider starting
out with a state consisting of $N_{i}$ atoms uniformly distributed on
a square lattice of volume $V$ and having a magnetization
corresponding to point F on the phase diagram. This state can lower
its free energy by forming a solid island with vapour around it, the
solid island being at point q of the phase diagram and the vapour at
point p, at the same external field $h_{f}$ as the solid.  One can
compute the change in the free energy in nucleating an island of up
spins of radius $r$ and this change is again composed of three pieces:
\newline
a)An increase in surface free energy given by 
\begin{equation}
\Delta G_{\text{edge}} = 2 \pi r \gamma,
\end{equation}
where $\gamma$ is the line tension or edge free energy per unit length
of the island.
\newline
b)The change in free energy in the region of the island that
condenses out.  This change is computed by taking the difference in
free energy between the initial state F and final state q and is given
by
\begin{equation}
\Delta G_{\text{c}} = \pi r^{2} (f^{>}[h_{f}] - f^{<}[h_{i}] + h_{f}m^{>}[h_{f}] - h_{i}m^{<}[h_{i}]).
\end{equation}
\newline
c) The change in free energy of the remaining region of volume
($V - \pi r^{2}$), as it moves from point F of the metastable part of the
phase diagram to point p, 
\begin{equation}
\Delta G_{\text{nc}} = (V - \pi r^{2})(f^{<}[h_{f}] - f^{<}[h_{i}] + h_{f}m^{<}[h_{f}] - h_{i}m^{<}[h_{i}]).
\end{equation}

The total change in free energy is thus
\begin{equation}
\Delta G_{\text{tot}} = 2 \pi r \gamma + 
\pi r^{2} (f^{>}[h_{f}] - f^{<}[h_{i}] + h_{f}m^{>}[h_{f}] - h_{i}m^{<}[h_{i}]) + 
(V - \pi r^{2})(f^{<}[h_{f}] - f^{<}[h_{i}] + h_{f}m^{<}[h_{f}] - h_{i}m^{<}[h_{i}]).
\label{lgt}
\end{equation}

Note that although the above equation for the free energy makes
it look like a function of two independent variables, $r$ and $h_{f}$,
there is only one independent variable. The second
variable is fixed by the constraint of conservation which can be
expressed as
\begin{equation}
V m_{i} = \pi r^{2} m^{>}[h_{f}] + (V - \pi r^{2})m^{<}[h_{f}].
\label{cons}
\end{equation}

Thus $\Delta G_{tot}$ can be looked upon as a function of $r$ alone by
replacing the final external field $h_{f}$ that appears in
Eq. (\ref{lgt}) with the value obtained by formally solving for
$h_{f}$ as a function of $r$ from Eq. (\ref{cons}).  Extremizing
Eq. (\ref{lgt}) w.r.t $r$ yields the radius of the island in
equilibrium with the gas.  This gives us the analog of the
Gibbs-Thomson formula for the lattice gas system,
\begin{equation}
\gamma + r \left( f^{>}[h] - f^{<}[h_{f}] + h_{f}\left( m^{>}[h_{f}] - m^{<}[h_{f}] \right) \right) + \frac{\partial h_{f}}{\partial r}\left( \frac{\pi r^{2} h_{f} \chi^{>}[h_{f}] + \left( V - \pi r^{2} \right) h_{f} \chi^{<}[h_{f}]}{2 \pi}\right) = 0,
\label{isingt}
\end{equation}
where $\chi [h] \equiv \partial m/\partial h$ is the susceptibility
and $\partial h_{f}/\partial r$ can be determined from
Eq. (\ref{cons}). Instead of regarding the above equation as an
equation in $r$ we substitute for $r$ in terms of $h_{f}$ using the
constraint [Eq. (\ref{cons})].  This enables us to solve the above
equation for $h_{f}$ numerically after substituting the series
expansions for the free energy, magnetization and susceptibility.  We
use the first thirteen terms in the series expansion.  On finding the
equilibrium final external field $h_{f}$, for a given initial density
of atoms, the equilibrium radius of the island at the extremum of the
free energy can be obtained using the constraint [Eq. (\ref{cons})].
The final field also tells us the final magnetization outside the
island (point p on Fig. \ref{mvshfig}) and hence the density of
adatoms outside $\rho_{f} = (1 + m^{<}[h_{f}])/2$.  This gives us the
required relation between the radius $r$ of the island vs. density of
gas outside $\rho_{f}$, which we refer to as a corrected Gibbs-Thomson
equation.  The solid line in Fig. \ref{gtfig} represents the curve for
the corrected Gibbs-Thomson formula. It is clearly seen that the
corrected theory gives better agreement with the simulations than the
continuum theory, particularly for islands of very small radii ($r <
8$ or $1/r > 0.125$).  This leads us to believe that the approximation
of an ideal gas of adatoms around the island is the principal cause
for the break down of the classical Gibbs-Thomson formula at high
vapour densities.

\section{Stability of islands and the Thermodynamic limit}
\label{sec:stab}
\begin{figure}
\epsfxsize=8.6 cm
\epsffile{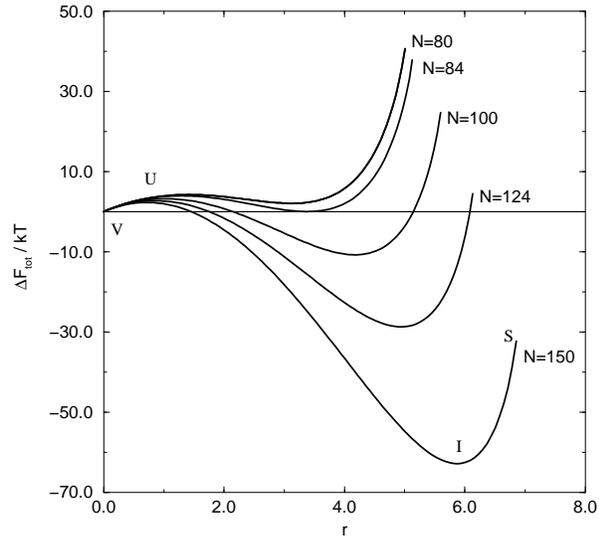}
\caption{The change in free energy as a function of r, 
for a system of volume V=10,000, for various values of N.  Notice the
global minimum of the Helmholtz free energy is a solid island of
radius r $\sim 5.9$, for the case $N=150$.  Further, if $N < 84$, the
globally stable extremum switches from island plus vapour (I) to pure
vapour (V)}.
\label{dffig}
\end{figure}

In this section we discuss the effects of finite size on the stability
of the islands we see in the simulation.  We first look at finite
size effects as predicted by the continuum version of the model we
have for a system of atoms (as in Section \ref{sec:gt}).  Figure
\ref{dffig} shows the effect of varying the number of atoms, $N$, at
constant volume $V$, on the total free energy change in nucleating an
island.  We see that the stable minimum (I) is no longer a global
minimum of the free energy of the system once $N$ falls below a
certain value and later this minimum vanishes completely (the curve
becomes flat) below a certain critical value of $N$ which we denote as
$N_{\text{cr}}(V)$, which evidently depends on $V$. Similar behaviour
is observed if we increase the volume $V$, at constant $N$. However if
we take the thermodynamic limit at constant initial density
($\rho_{i}$ = constant, $V \rightarrow \infty$) the stable minimum
persists and moves off towards $r = \infty$.  These results can be
understood by means of a stability analysis.

The equilibrium between an island and the vapour around it
is dynamic in nature and can be understood as a balance between the
rate at which atoms from the vapour attach themselves to the perimeter
of the island and the rate at which atoms detach themselves from the
perimeter of the island to become part of the vapour.  The former rate
would be proportional to the density of the vapour surrounding the
island while the latter would be governed purely by temperature and
would be independent of the density of vapour surrounding the
island, in the low density limit.

Consider a change in the radius of an island in
equilibrium with its vapour.  If the island grows from an initial
radius $r$ to a radius $r + dr$ by swallowing some atoms from the
vapour phase the concomitant change in the density of the vapour would
be
\begin{equation}
d\rho_{f} = - \frac{2 \pi  r (\rho_{s} - \rho_{f})dr}{V - \pi r^{2}}.
\label{dconst}
\end{equation}
If the new island of radius $r+dr$ is to be in equilibrium with vapour
around it, one can compute the change in equilibrium vapour density
around it (i.e., the difference between the vapour density around an
island of radius $r+dr$ and the vapour density around an island of
radius $r$) from the Gibbs-Thomson formula [Eq. (\ref{consgt})], 
\begin{equation}
d\rho_{f} = - \frac{\gamma \rho_{f}dr}{k T r^{2} (\rho_{s}-\rho_{f})}.
\label{dgt}
\end{equation}
The above two equations predict that the density will decrease if the
island grows ($dr > 0$) which is to be expected.  If the actual change
in density [Eq. (\ref{dconst})] is larger in magnitude (smaller in
value) than that dictated by equilibrium [Eq. (\ref{dgt})] the island
would be stable. This is because the new density around the island is
too low and consequently the new attachment rate would be lower than
the detachment rate.  This in turn would force more atoms to detach
from the island thus bringing down the size of the island.  From this
one can conclude that for stability one needs
\begin{equation}
- \frac{2 \pi  r (\rho_{s} - \rho_{f})dr}{V - \pi r^{2}} <
- \frac{\gamma \rho_{f}dr}{k T r^{2} (\rho_{s}-\rho_{f})}  ,
\end{equation}
which can be written as
\begin{equation} 
r^{3} > \frac{\gamma \rho_{i} V}{2 \pi k T(\rho_{s} -
\rho_{i})^{2}}\left(1 - \frac{\pi r^{2}}{V}
\right)^{2}\left(1 - \frac{\rho_{s} \pi r ^{2}}{\rho_{i} V} \right) ,
\label{stabcond}
\end{equation}
where $\rho_{i} = N/V$ as before.  We see from this that for stability
the radius of an island should be greater than a certain minimum value
which is obtained by solving
\begin{equation}
r_{min} =  \left(\frac{\gamma \rho_{i} V}{2 \pi k T(\rho_{s} -
\rho_{i})^{2}}\right)^{1/3} \left(1 - \frac{\pi r^{2}_{min}}{V}
\right)^{2/3}
\left(1 - \frac{\rho_{s} \pi r^{2}_{min}}{\rho_{i} V} \right)^{1/3}.
\label{rmin}
\end{equation}
Along with this if we use the Gibbs-Thomson formula [Eq. (\ref{gt})]
we can obtain a relation for the critical value $N_{cr}$ as a function
of the volume.  All issues of local stability of the islands can be
resolved using these equations. The curve $N_{cr}(V)$ in $N-V$ space
defines a boundary between regions where one can have stable islands
and regions where one can have no stable islands \cite{foot3}. 

One can show that for large system size the last two terms in the
product of Eq. (\ref{rmin}) go to unity and we have
\begin{equation}
r_{min}^{3} = \gamma \rho_{i} V/\left[2\pi kT(\rho_{s}-\rho_{i})^{2}\right].
\end{equation}
This shows that the minimum radius of a stable island grows as the
one third power of the volume of the box in two dimensions.

We now digress to note the behaviour of the unstable root (U) of the
free energy in Fig. \ref{dffig} in the thermodynamic limit. It is seen
that the unstable root does not scale with system size by plotting
this root obtained by numerical solutions versus the system volume V.
The unstable root reaches a limiting value in the limit $V \rightarrow
\infty$, which can be obtained from Eq. (\ref{consgt}) by neglecting
terms of order $r^{2}/V$. The critical radius $r^{\star}$, which is
obtained by taking this limit is given by
\begin{equation}
r^{\star} = \frac{\gamma}{k T \left[\rho_{s} \ln
\left(\rho_{i}/\rho_{\infty} \right) +
\left( \rho_{\infty} - \rho_{i} \right)\right]}.
\end{equation}
This form is identical to the form for the critical radius
quoted in the context of nucleation theory \cite{abr}. The nucleation
barrier, which is the free energy barrier the system of supersaturated
vapour should overcome in order to form a stable island plus vapour,
attains a limiting value of
\begin{equation}
\Delta F^{\star} = \frac{\pi \gamma^{2} \left[\rho_{s} \ln \left(\rho_{i}/\rho_{\infty}
\right) + \rho_{\infty} + \rho_{s} - 2 \rho_{i} \right]}{k T \left[ \rho_{s} \ln \left( 
\rho_{i}/\rho_{\infty} \right) + \rho_{\infty} - \rho_{i} \right]^{2}}.
\end{equation}

\begin{figure}
\epsfxsize=8.6 cm
\epsffile{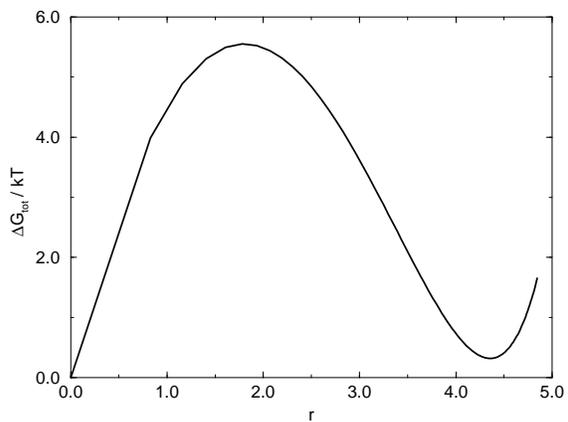}
\caption{The free energy of an island of radius r 
plotted for a system of 109 particles with V = 10,000}
\label{unstablefig}
\end{figure}

How about seeing the unstable islands in our simulation?  We have
observed that if we start out with 109 atoms in a 100x100 system at a
temperature of 1347 K, the island size fluctuates considerably and
there are several frames of data where the island breaks up into many
smaller ones.  This can be understood within the framework of our
theory for the Ising model.  Fig. \ref{unstablefig} shows the change
in free energy on nucleating an island of radius $r$ in the Ising
model for 109 particles.  From this we see that the island-vapour
system is not a point of global minimum of free energy.  Further the
nucleation barrier to go from this state to one of uniform vapour is
given by $\Delta G/kT = 5.26$. Also we can see from this figure that
the fluctuations to various other island sizes are not highly
unlikely.  This would account for the large fluctuations in island
size.  The same effect is seen for 25 particles at a temperature of
1000 K.

\section{Investigation of Microscopic Origins}
\label{sec:mic}
Since these simulations of atomic scale systems exhibit the
Gibbs-Thomson effect viz. an enhanced vapour pressure around islands
of small radii relative to the vapour pressure outside a flat
interface, the opportunity arises to investigate the relationship
between this thermodynamic effect and the microscopic dynamics.  We
may ask, from a microscopic point of view, what is the origin of the
enhanced adatom vapor concentration in equilibrium with a small
island.  A complete discussion of this issue involves many details of
the microscopic characteristics of the island-vapor interface, which
are beyond the scope of this paper.  Here we outline our main
findings; the interested reader is referred to Refs. \onlinecite{mclean96a} 
and \onlinecite{mcleanprep} for further details.

\begin{figure}
\epsfxsize=8.6 cm
\epsffile{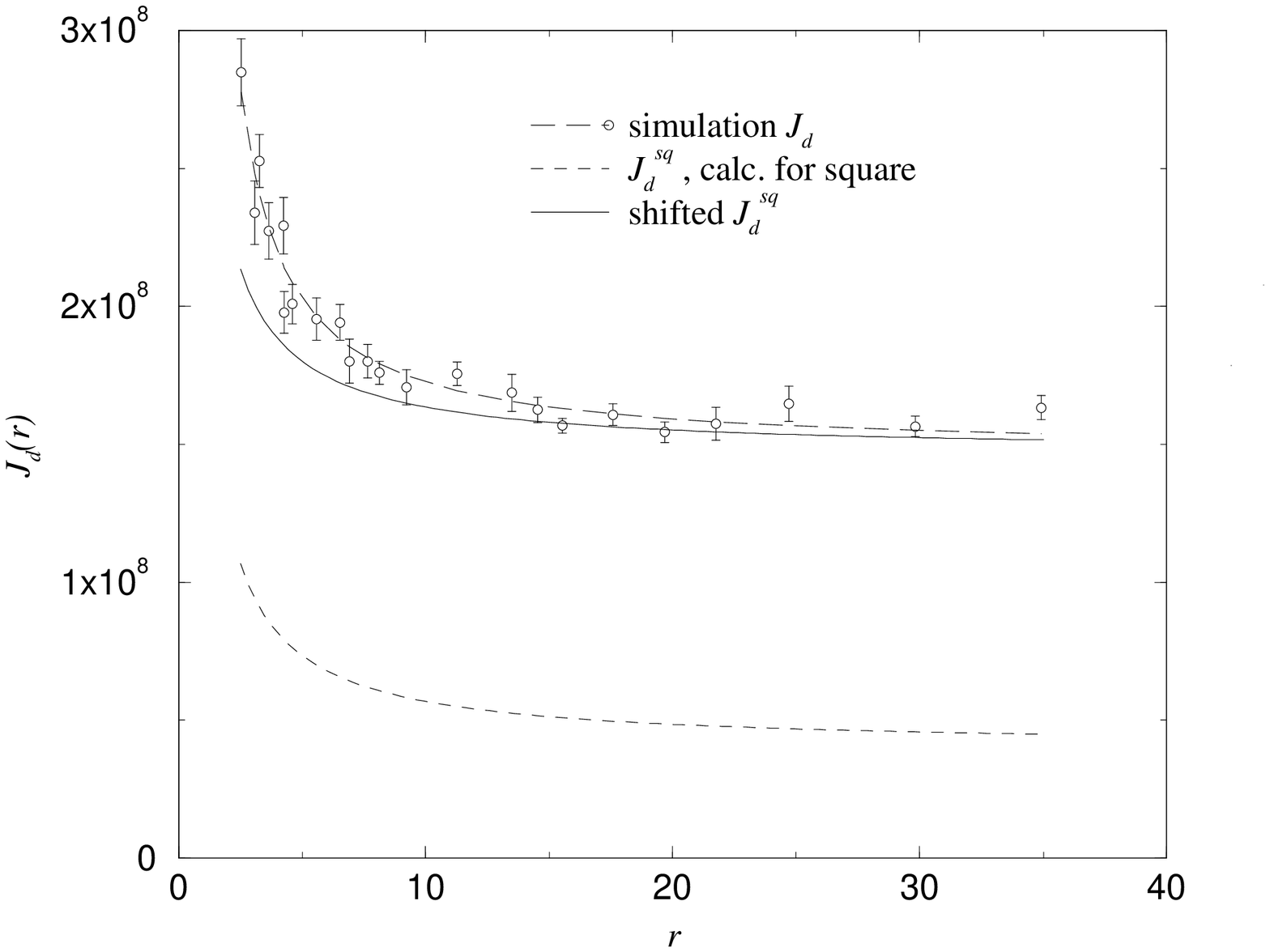}
\caption{A comparison of the changes in detachment current density
found for islands in simulation ($J_d$),
and for a square island treated in the same way ($J_d^{sq}$).
The solid line represents $J_d^{sq} +A$, where $A$ is chosen
such that it $J_d-(J_d^{sq} +A) \to 0$ as $r\to\infty$.\protect ~
Lines, to guide the eye, are fits to the form $f(\infty)\exp(C/r)$.}
\label{fig:micros}
\end{figure}

As discussed in Section~\ref{sec:stab}, equilibrium between the island
and vapour implies detailed balance at the interface: atoms are
attaching to and detaching from the island with equal rates.  Analysis
of our simulations shows that for small equilibrium islands, the
interface transfer activity is enhanced in proportion to the vapour
density.  For example, the data points in Fig. \ref{fig:micros} show
the rate at which atoms detach from an island per unit length of the
macroscopic island-vapor interface.  This leads to the following
microscopic interpretation of the Gibbs-Thomson effect.  As the island
size decreases, it becomes easier for atoms to detach from it, raising
the detachment current density.  However, we find that there is no
noticeable change in the ease with which an atom can attach to the
island for islands whose radii vary between 4 and 35.  Therefore, a
higher vapour density is required to maintain dynamic equilibrium.

The enhanced detachment current density for smaller islands can be
ascribed to trends in the character of the sites on the island edge.
On smaller islands, the density of edge atoms is found to actually
decrease, so that there are fewer atoms per unit length of interface
available for detachment. However, the average coordination of the
edge atoms is found to be smaller, which leads to lower energy
barriers for edge atom motion.  Also, each edge atom on a smaller
island tends to have more detachment moves available to it.  That is,
when the edge atom moves it is more likely to detach, as opposed to
moving along the edge of the island.  The net result of these trends
yields the observed enhancement in detachment.

\begin{figure}
\epsfxsize=8.6 cm
\epsffile{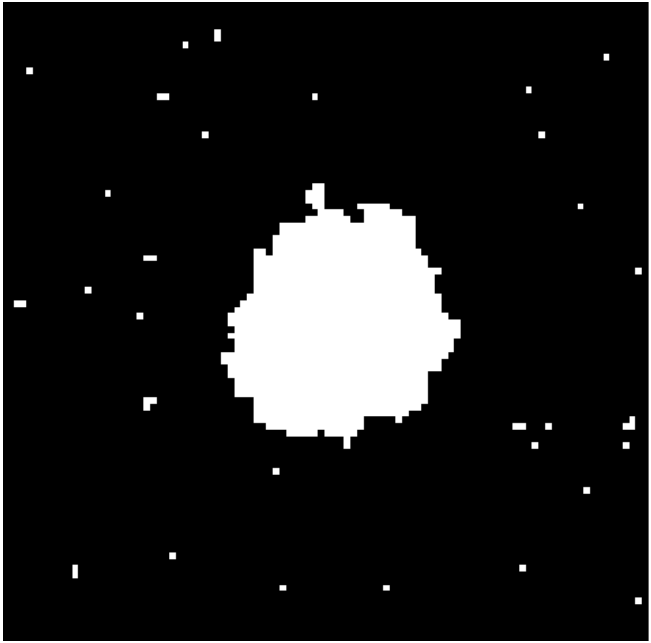}
\caption{A snapshot of an island with vapour around it as 
seen in the simulation}
\label{pictfig}
\end{figure}

Note that the above trends, observed in the equilibrium islands (e.g.,
Fig. \ref{pictfig}) of our simulations also hold true for a square
island, although a square is not the thermodynamic shape of an
equilibrium island at finite temperatures.  As a square island is made
smaller, the corner sites acquire greater significance.  Since the
corner sites of a square island have a lower coordination than sites
on the side of a square, the average coordination of edge atoms on a
small square is lower than it is on a large square.  Similarly, corner
atoms have two detachment moves available, while side atoms have only
one.  Therefore a smaller square has a higher ratio of available
detachment moves to number of edge atoms.

This analogy between the simulated islands and square islands suggests
that an important element of the observed behavior is the simple
geometric constraint that any closed perimeter on a square lattice
must have four more outward pointing corners than it has inward
pointing crevices.  As a test of this idea, Fig. \ref{fig:micros}
compares the detachment current density observed in the simulation
with that expected for a square island of the same area and at the
same temperature.  As expected, the overall detachment current density
is lower for the square island, as it has the smoother edge.  However,
as the island size is varied, the magnitude of the enhancement in
detachment from the square is comparable to the enhancement in
detachment from the simulated islands.  It is therefore clear that it
is important to consider the effect of the `four extra corners' in an
understanding of the Gibbs-Thomson effect at a microscopic level.  It
is difficult to quantify the effect of this geometrical constraint, as
it is impossible to label an individual corner on an equilibrium
island as being due to either geometry or thermal roughening.
However, comparison with the non-equilibrium square island gives an
indication of the strength of the effect.

\section{Conclusions}
\label{sec:conc}
We have simulated a lattice gas to mimic the behaviour of a cluster of
atoms, on generic surfaces, in an effort to study the relationship
between the cluster radius and the vapour density around it.  We have
shown that the ``classical'' Gibbs-Thomson relationship one computes
assuming an ideal gas of atoms is incorrect at high vapour densities
and a knowledge of the true equation of state is necessary to obtain a
better result.  We have seen that the corrected formula can be used
down to islands with about 150 atoms at a temperature of 0.6$T_{c}$
and islands with about 30 atoms at 0.445$T_{c}$, in the case of our
simulations.

Further we have seen how metastable states in traditional nucleation
theories can be made stable by finite size effects.  We have seen how
these states may arise in the context of the Ising model and have
explored the metastable continuation of the equation of state in the
Ising model. Simulations performed on the Ising model agree well with
our predictions regarding stability.

As far as experimental observations of the corrections to the
Gibbs-Thomson formula are concerned such an effect would surely be
observed in a system with short range interactions at small island
sizes and high temperature (about 60 percent of the melting
temperature).  However in real situations in addition to the short
range attractive forces that bind atoms to each other there exist long
range dipolar forces at step edges, between the atoms at the edge and
the vapour. This may skew the predictions of a theory like ours which
is simple and ignores such effects.  Finally we have looked at the
microscopic origins of the Gibbs-Thomson formula and have offered
heuristic arguments that it maybe correlated to geometric constraints.

\section{Acknowledgments}
BK would like to thank Gerard Barkema for helping out with a lot of
information regarding the MC method and for the data on energy
barriers for which we are grateful to Rien Breeman.  He would also
like to thank Mark Newman for help on series expansions. We would
like to thank Eric Chason for a lot of help on details regarding the
simulation. BK wishes to acknowledge financial support from the
National Science Foundation and the Materials Science Centre through
grants NSF-GER-9022961 and DMR-91-21654 respectively, while JM
acknowledges support from the Air Force Office of Sponsored Research
(grants AFOSR-91-0137 and AFOSR/AASERT F49620-93-1-0504) and partial
support from the Cornell Materials Science Centre (NSF-DMR-91-21654).
This work made use of the MSC Multi-User-Computer Facility, an MRL
Central Facility supported by the National Science Foundation under
Award No. DMR-9121564.

\begin{table}
\caption{Energy barriers for intra-layer atomic moves}
\label{bartab}
\begin{tabular}{cdddd}
\bf{Initial}      &\multicolumn{4}{c}{\bf Final Coordination} \\ \cline{2-5}
\bf{Coordination} & 0-fold   & 1-fold   & 2-fold   & 3-fold   \\ \hline
0-fold & 0.697 eV & 0.479 eV & 0.328 eV & 0.166 eV \\ 1-fold & 0.820
eV & 0.624 eV & 0.450 eV & 0.275 eV \\ 2-fold & 1.010 eV & 0.791 eV &
0.591 eV & 0.377 eV \\ 3-fold & 1.189 eV & 0.957 eV & 0.718 eV & 0.462
eV \\
\end{tabular}
\end{table}

\begin{table}
\caption{Constants for the Ising model for bond energy = 0.341 eV}
\label{isingtab}
\begin{tabular}[c]{lllll}
&\multicolumn{2}{c}{T=1347 K}&\multicolumn{2}{c}{T=1000 K} \\
\hline
$T_{c}$           & 2245      &  K   & 2245      & K      \\
$\gamma_{\text{min}}$    & 0.1161    &  eV  & 0.1465    & eV     \\
$\gamma_{\text{max}}$    & 0.1184    &  eV  & 0.1543    & eV     \\
$\gamma_{\text{avg}}$    & 0.1173    &  eV  & 0.1507    & eV     \\
$\rho_{\infty}$   & 0.003578  &      & 0.000396  &        \\
$\rho_{s}$        & 0.996422  &      & 0.999602  &        \\
\end{tabular}
\end{table}

\begin{figure}
\epsfxsize=8.6 cm
\epsffile{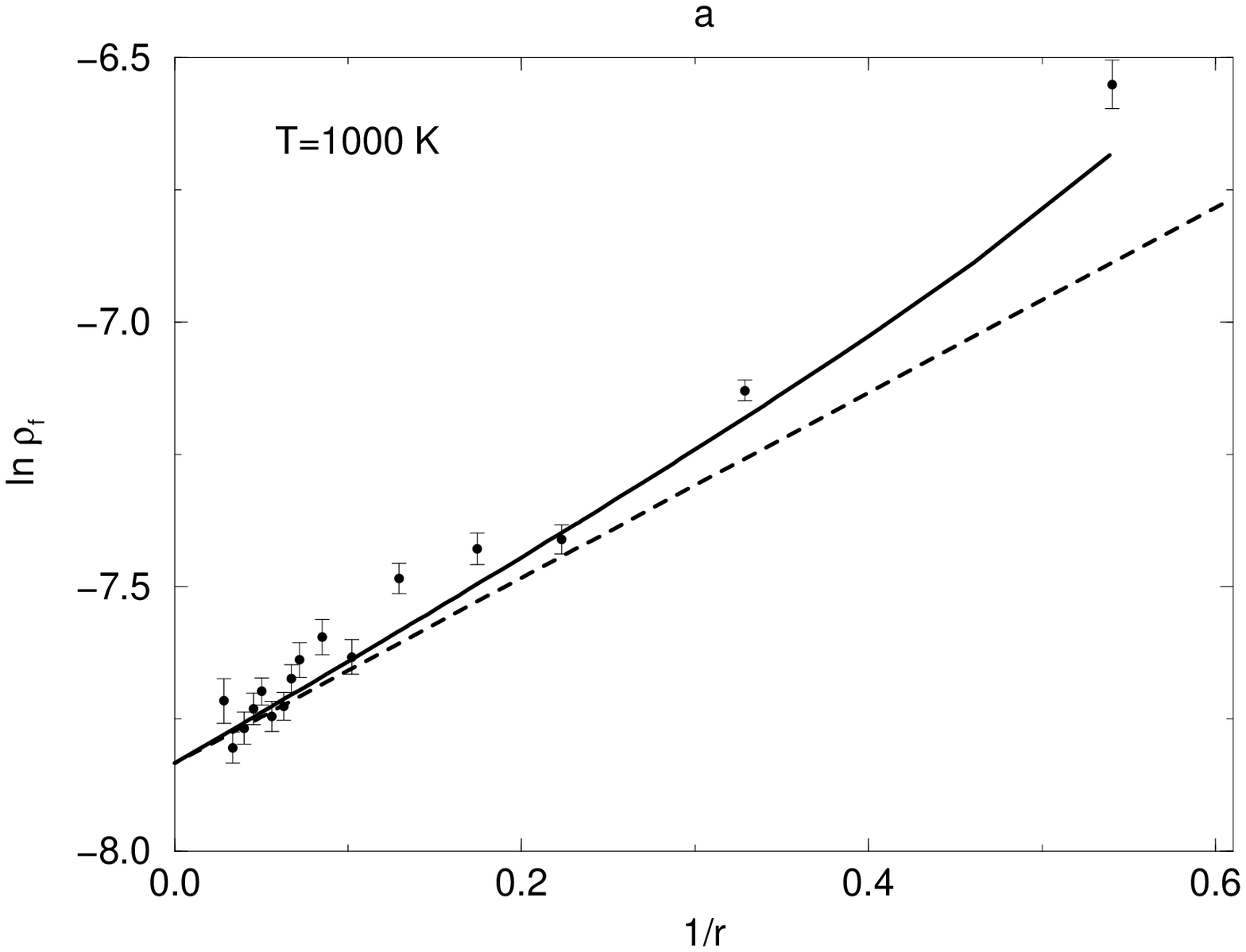}
\epsfxsize=8.6 cm
\epsffile{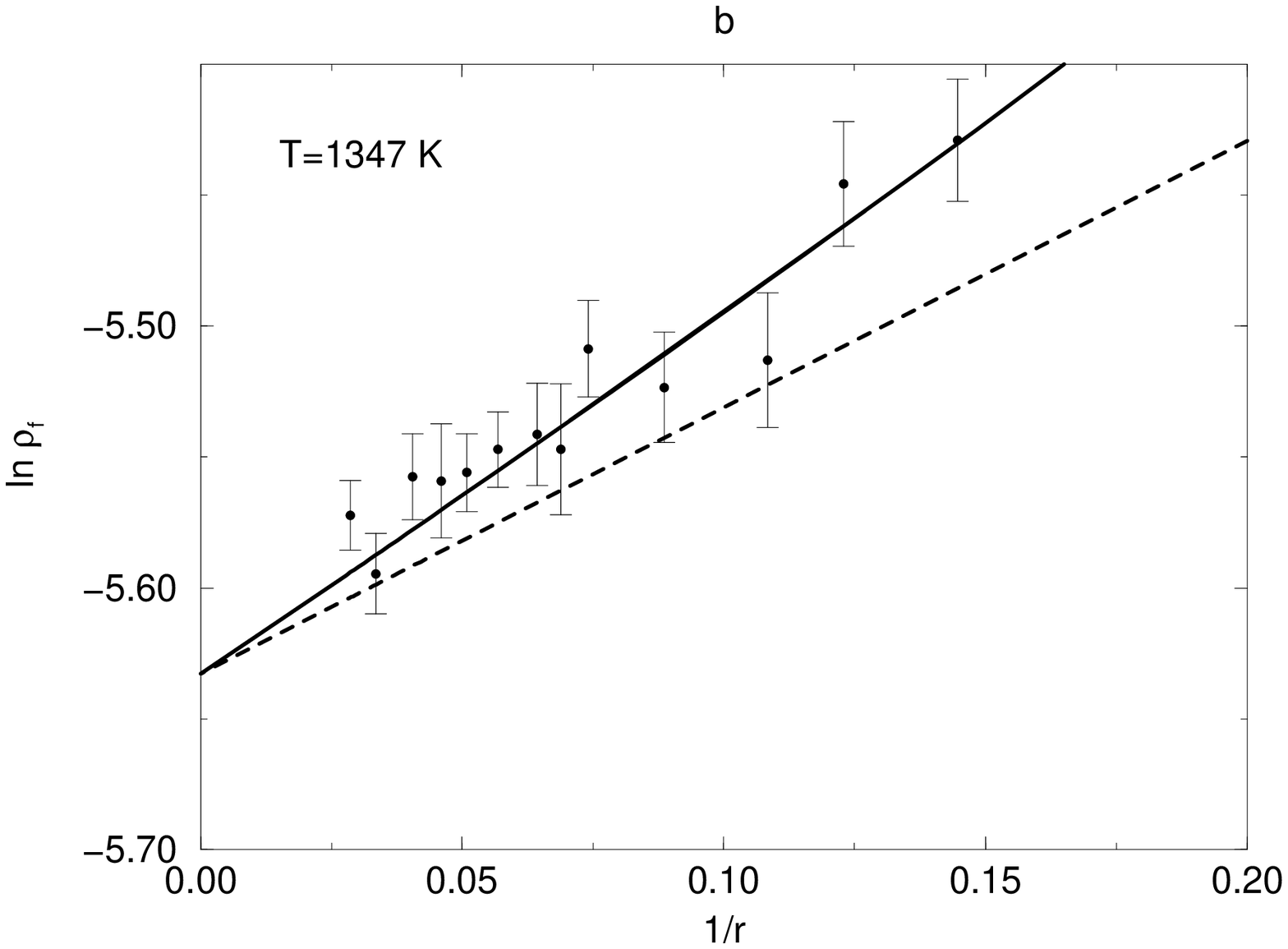}
\caption{Plot of the logarithm of the density of vapour outside an island vs. 
the reciprocal of its equilibrium radius.  The dashed line represents
the Gibbs-Thomson prediction assuming an ideal gas of vapour.  The
solid curve is the prediction using the corrected Gibbs-Thomson formula
for the Ising model.  Fig. a is the data at a temperature of 1000 K
while Fig. b is at 1347 K}
\label{gtfig}
\end{figure}

\begin{references}
\bibitem{foot1}Note that the solid island (point 5) is not at the same
pressure as the vapour in equilibrium around it (point 2) if the
island has a finite radius of curvature.  The difference in pressure
between the solid and gas is given by the Young-Laplace equation in
two dimensions, $\Delta p = \gamma/r$, where $\gamma$ is the edge free
energy and $r$ the radius of curvature of the island.
\bibitem{foot2}One may suspect the assumption of zero compressibility 
for the solid phase may be the reason for the discrepancy between the
data and the fit using the Gibbs-Thompson formula, as opposed to
``non-ideal'' behaviour of the gas around the island.  However this is
ruled out because we have rederived the Gibbs-Thomson formula,
assuming the vacancies in the solid phase to behave as an ideal
gas. The curve so obtained is virtually indistinguishable from the
dotted curve in Fig. \ref{gtfig} which represents the classical
uncorrected Gibbs-Thomson formula.  The inclusion of the extra term on
the right hand side of Eq. (\ref{consgt}) does not produce any
discernable change either to the classical uncorrected Gibbs-Thomson
curve.
\bibitem{foot3} An alternate route to obtaining these relations 
would be to make sure that we are at a local minimum of the total
change in free energy that we derived in Section \ref{sec:gt},
Eq. \ref{dftot}, i.e.,
\begin{equation}
\frac{d \Delta F_{tot}}{dr} = 0, 
\label{gthoms}
\end{equation}
\begin{equation}
\frac{d^{2}\Delta F_{tot}}{dr^{2}} \geq 0.
\label{dfminimum}
\end{equation}
The first of the above two equations is just the Gibbs-Thomson formula
as we have seen in Section \ref{sec:gt} while the second one 
which expresses the fact that we be at a local minimum of the
free energy, reduces to Eq. \ref{stabcond}.
\end{references}
\end{document}